\documentclass{PoS}

\title{Search for Gauginos and Gauge Mediated SUSY Breaking Scenarios at LEP}

\ShortTitle{Search for Gauginos and Gauge Mediated SUSY Breaking Scenarios at LEP}

\author{\speaker{Gabriella P\'asztor}\thanks{
On leave of absence from KFKI RMKI, Budapest.}\\
University of California, Riverside\\
E-mail: \email{Gabriella.Pasztor@cern.ch}
}

\abstract{Data collected by the OPAL detector at LEP up to the highest 
energies of 209 GeV are analyzed to search for evidence of supersymmetric
(SUSY) particle production. \\
Chargino pair-production (e$^+$e$^- \rightarrow \tilde\chi_1^+\tilde\chi_1^-$)
and neutralino associated production (e$^+$e$^- \rightarrow
\tilde\chi_1^0\tilde\chi_2^0$) are considered in models where supersymmetry
breaking is mediated by gravitational interactions from the hidden sector to
the visible sector of the Standard Model and SUSY particles. R-parity
conservation is assumed. The focus of the searches is on topologies arising
when scalar fermions are heavy and the lightest neutralino is the lightest SUSY
particle (LSP). \\
Searches for topologies predicted by gauge-mediated SUSY breaking (GMSB) models
are also discussed. In these models the LSP is the gravitino and the
phenomenology is driven by the nature of the next-to-lightest SUSY particle
(NLSP) which is either the lightest  neutralino, stau or mass degenerate
sleptons. As the NLSP decay length is effectively unconstrained, all possible
values are considered and results independent of the NLSP lifetime are
presented for  all relevant final states including direct NLSP pair-production
and NLSP production via cascade decays of heavier SUSY particles.
\\
None of the searches shows evidence for SUSY particle production. Cross-section
limits are presented at the 95\% confidence level both for direct NLSP
production and, for the first time, also for cascade decays,  providing the most
general, almost model independent results. These are then interpreted in the
framework of minimal theoretical  models.  Large areas of the accessible
parameter space are excluded. In the super-gravity inspired Constrained MSSM
(CMSSM) we derive the following limits for $m_0>500$~GeV, $A_0=0$ and 
$\Delta M > 5$~GeV  on the lightest gaugino masses:  
$m_{\tilde\chi_1^\pm} > 101$~GeV, $m_{\tilde\chi_1^0} > 40$~GeV  and
$m_{\tilde\chi_2^0}> 78$~GeV, independent of $\tan\beta$. 
In the minimal GMSB model, the NLSP masses are constrained to be
$m_{\tilde\chi_1^0} > 53.5$~GeV, $m_{\tilde\tau_1} > 87.4$~GeV and 
$m_{\tilde\ell} > 91.9$~GeV in the neutralino, stau and slepton co-NLSP
scenarios, respectively.}

\FullConference{International Europhysics Conference on High Energy Physics\\
		 July 21st - 27th 2005\\
		 Lisboa, Portugal}

\begin{document}

\section{Introduction}

Supersymmetry, the best proposed solution to the problems of the Standard Model
(SM), postulates the existence of a partner for each SM particle chirality
state. The discovery of these superparners would be the most direct evidence
for SUSY. Since SUSY particles are not observed with the same mass as their SM
partners, SUSY must be broken. In the most widely investigated scenarios, it is
assumed that SUSY is broken in some {\em hidden} sector of new particles and is
{\em communicated} to the {\em visible} sector of SM and SUSY particles by
gravity or gauge interactions. 

We review the results of the searches for chargino and neutralino
production~\cite{opal_chargino} motivated by gravity-mediated SUSY breaking
models, and then present a study of gauge-mediated SUSY breaking
topologies~\cite{opal_gmsb}  using the data collected by the OPAL detector at
LEP up to the highest center-of-mass energies of 209 GeV.

\section{Searches for chargino and neutralino production}

The lightest charginos are expected to be pair-produced at LEP with  a
cross-section of typically a few pb. However, if  the electron sneutrino is
light, the cross-section decreases due to destructive interference between
$s$-channel $\gamma$ / Z and $t$-channel $\tilde\nu_\mathrm{e}$ exchanges.  If
charginos are heavy, neutralino associated production may be the only SUSY
signal at LEP.

The focus of the analyses~\cite{opal_chargino} 
is on topologies arising when scalar fermions are
heavy and the lightest neutralino is the LSP:
$\mathrm{e}^+\mathrm{e}^- 
\rightarrow 
\tilde\chi^+_1 \tilde\chi^-_1 
\rightarrow 
\mathrm{W}^{+\ast} \tilde\chi^0_1 \mathrm{W}^{-\ast} \tilde\chi^0_1$
%$\rightarrow
%\ell^+ \nu_\ell \tilde\chi^0_1 
%\ell^- \overline\nu_\ell \tilde\chi^0_1$ 
%(leptonic),
%$\mathrm{q} \overline\mathrm{q} \tilde\chi^0_1
%\ell \nu_\ell \tilde\chi^0_1$
%(semi-leptonic),
%$\mathrm{q} \overline\mathrm{q} \tilde\chi^0_1
%\mathrm{q} \overline\mathrm{q} \tilde\chi^0_1 $
%(hadronic), 
and 
$\mathrm{e}^+\mathrm{e}^- \rightarrow \tilde\chi^0_2\tilde\chi^0_1 
\rightarrow 
(\mathrm{Z}^{0\ast} \tilde\chi^0_1) (\tilde\chi^0_1)$.
%$\rightarrow
%\mathrm{q}\overline\mathrm{q} \tilde\chi^0_1 \tilde\chi^0_1$.
The kinematics strongly depend on the mass difference $\Delta M$ between
$\tilde\chi_1^\pm$ (or $\tilde\chi_2^0$) and $\tilde\chi_1^0$. 

Despite a small general excess in some of the channels, there is no indication
for chargino or neutralino pair-production in the OPAL data. Upper
cross-section limits are derived: 0.1 pb is achieved almost up to the kinematic
limit for leptonic and 0.3 pb for semi-leptonic and hadronic final states in 
the chargino search, while 0.1 pb for the hadronic final state of the 
neutralino search. 

\begin{figure}
\vspace*{-0.5cm}

\includegraphics[width=0.33\textwidth, height=0.3\textwidth]{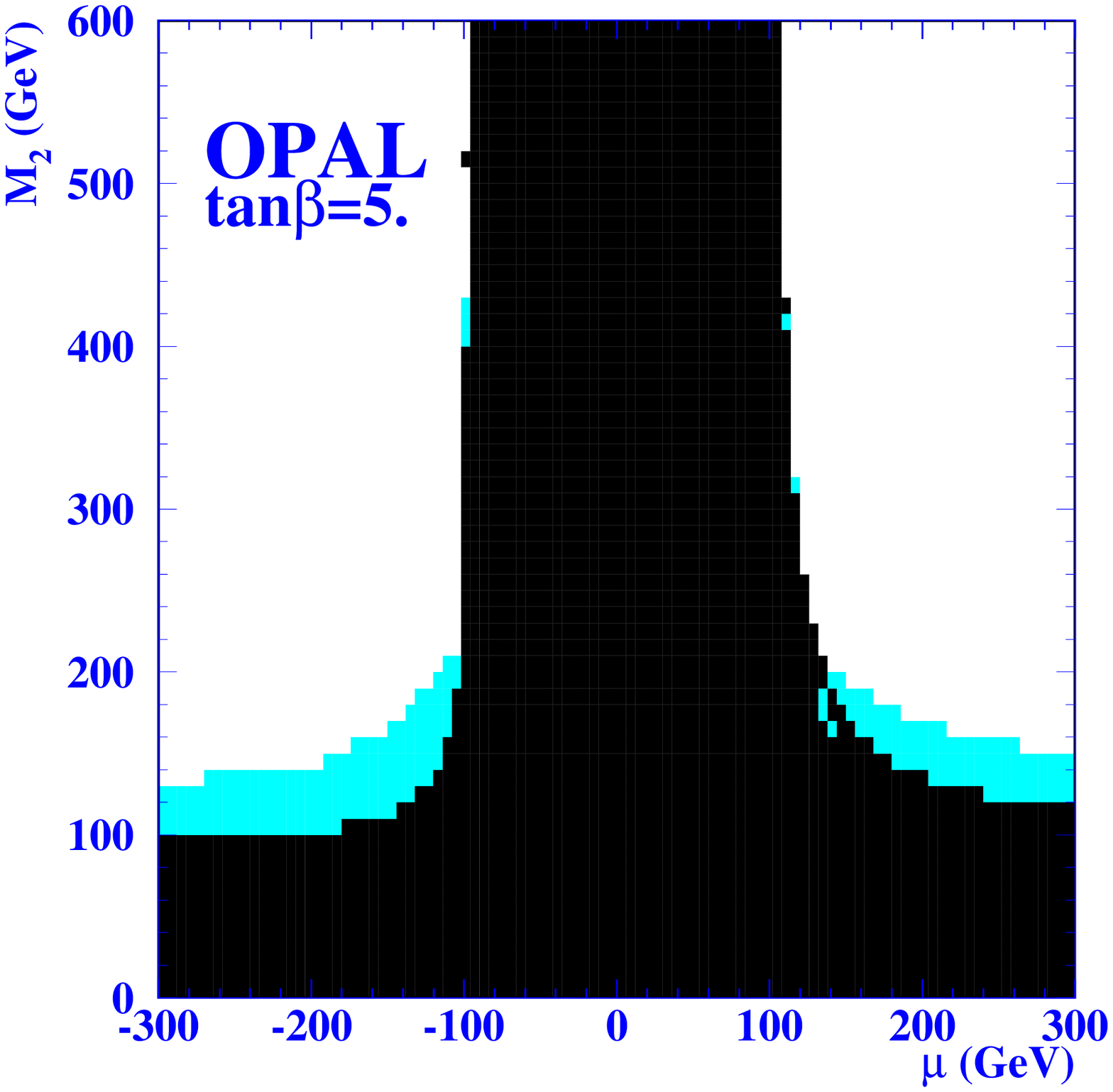}
\includegraphics[width=0.33\textwidth, height=0.3\textwidth]{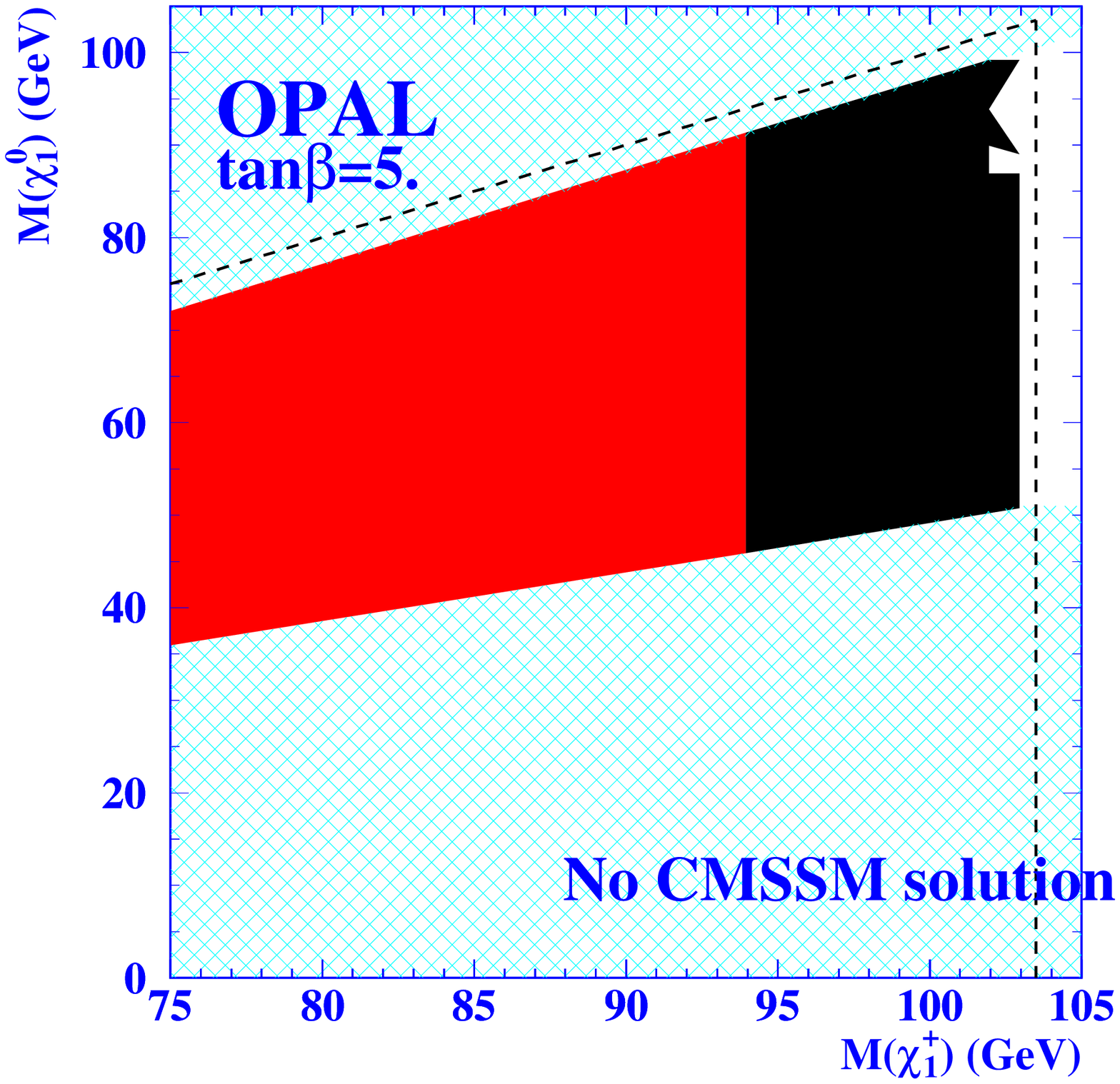}
\includegraphics[width=0.33\textwidth, height=0.3\textwidth]{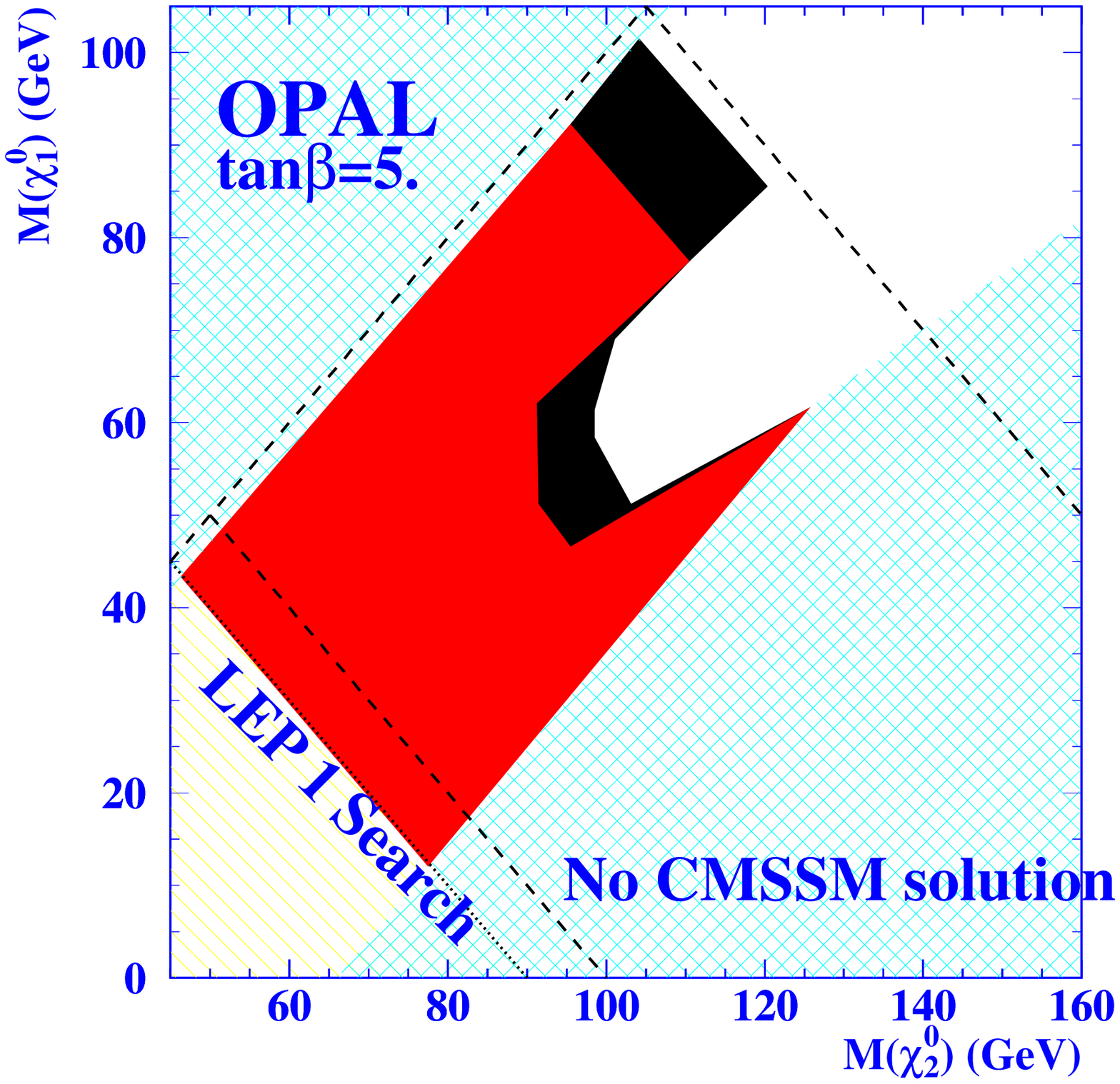} 
\vspace*{-0.8cm}

\caption{({\it left}) Excluded region of the CMSSM parameter space (black) together
with the kinematically allowed region (light blue/grey). ({\it center}) Excluded
regions in the $\tilde\chi_1^\pm - \tilde\chi_1^0$ mass plane by the 189 GeV
(red/grey) and the highest energy (black) OPAL data. ({\it right}) Excluded region in
the  $\tilde\chi_2^0 - \tilde\chi_1^0$ mass plane.}
\label{fig:gaugino}
\end{figure}

The results are interpreted in CMSSM where all gauginos are assumed to have  a
common mass ($m_{1/2}$) at the GUT scale, implying a relation between their
masses ($M_i$) at the electroweak scale, and all sfermions to have a common
mass ($m_0$) at the SUSY breaking scale. Further parameters are: the
pseudo-scalar Higgs boson mass ($m_\mathrm{A}$), the ratio of the v.e.v. of the
two Higgs field doublets ($\tan\beta$), the mass mixing parameter of the Higgs
field ($\mu$) and the trilinear sfermion-Higgs coupling at the GUT scale
($A_0$). As gaugino properties depend mainly on $M_2, \mu$ and $\tan\beta$,
these parameters are scanned in detail, while the parameters determining the
sfermion masses are fixed to $m_0=500$~GeV and $A_0=0$. The exclusion in the
$\mu - M_2$ plane is shown for $\tan\beta=5$ on Figure~\ref{fig:gaugino}.
The strongest exclusions are found for large $\tan\beta$. These exclusions can
be turned into constraints on the masses as illustrated on
Figure~\ref{fig:gaugino}. For $\Delta M>5$~GeV, we derive lower limits of
101, 40 and 78 GeV  on the masses of $\tilde\chi_1^\pm$, $\tilde\chi_1^0$ and
$\tilde\chi_2^0$, respectively, independent of $\tan\beta$.

\section{Searches for GMSB topologies}

The main new feature of models with GMSB is a light gravitino LSP. The
phenomenology is driven by the nature of the NLSP, which is either the 
lightest neutralino, stau or  mass-degenerate sleptons. As the gravitino
couples very weakly to heavier SUSY particles, those will decay typically to
the NLSP which then decays via  $\tilde\chi^0_1 \rightarrow \gamma
\tilde\mathrm{G}$ or  $\tilde\ell \rightarrow \ell \tilde\mathrm{G}$. We
study~\cite{opal_gmsb}
all relevant final states: direct NLSP production and  its appearance in the
decay chain of heavier SUSY particles, like charginos, neutralinos and
sleptons.  

\begin{figure}
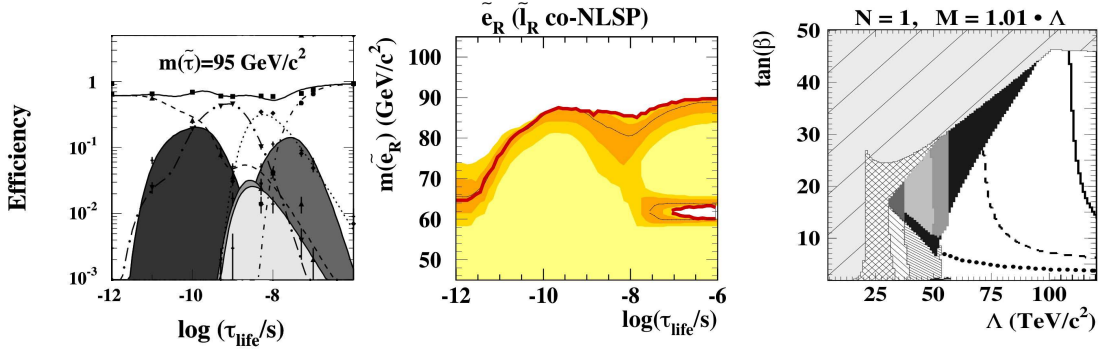

\vspace*{-0.4cm}

\includegraphics[width=0.095\textwidth]{pr409_06.epsi.y.epsf}
\includegraphics[width=0.22\textwidth, height=0.27\textwidth]{pr409_06.epsi.epsf}
\includegraphics[width=0.325\textwidth, height=0.3\textwidth,]{pr409_19.epsi.epsf}
\includegraphics[width=0.325\textwidth, height=0.3\textwidth,]{pr409_20.epsi.epsf}
\vspace*{-0.8cm}

\hspace*{1.3cm}\includegraphics[width=0.182\textwidth]{pr409_06.epsi.x.epsf}
\vspace*{-0.4cm}

\caption{({\it left}) Efficiencies for stau pair-production at
$\sqrt{s}=208$~GeV. The symbols represent the
efficiencies for ten simulated lifetimes while the curves show the interpolating
efficiency functions of the searches for promptly decaying staus (dashed), large
impact parameters (long dash-dotted), kinks (dotted) and stable staus
(dash-dotted) together with the overlap efficiencies (filled areas). The total
efficiency is shown by full line. ({\it center}) Observed (thick red/dark grey) and 
expected (thin black) lower mass limits for pair-produced selectrons in the
slepton co-NLSP scenario. The 68\% and 95\% probability intervals are shown by
orange/grey shades. ({\it right}) Regions in the $\Lambda - \tan\beta$ plane excluded by 
the different searches in the stau and slepton co-NLSP scenarios (direct
NLSP production - black, chargino - dark grey, neutralino - grey, selectron and
smuon production with stau NLSP - light grey) and in the neutralino NLSP scenario
(neutralino production - dense hatched, chargino - hatched). The 
LEP1 search regions is cross-hatched and the theoretically not allowed region is
grey with sparse hatching. The constraint from the LEP combined Higgs limit of
$m_\mathrm{H}>114.4$~GeV is also indicated (full line) together with large 
effect weakening it from theoretical (dashed) plus top quark mass (dotted) 
uncertainities.}
\label{fig:gmsb}
\end{figure}

The minimal GMSB model introduces five new parameters and a sign:  the SUSY
breaking scale ($\sqrt{F}$), the SUSY particle mass scale ($\Lambda$), the
messenger mass ($M$), the number of messenger sets ($N$), $\tan\beta$ and
sign($\mu$). As the decay length of the NLSP depends on $\sqrt{F}$ and is
effectively unconstrained, NLSP decays inside and outside of the detector are
searched for. With increasing decay length, the event signatures include: 
energetic leptons or photons and  missing energy due to the undetected
gravitino, tracks with large impact parameters, kinked tracks, or heavy
long-lived charged particles. In total more than 14 different selections are
developed to cover the GMSB topologies. The results are
combined (with special attention to treat the overlaps among the many channels
properly) to get lifetime independent results. To achieve a good description of
the selection efficiencies over the whole mass and lifetime range at all
center-of-mass energies, without generating an excessive number of Monte Carlo
samples, an interpolating function is determined. On Fig~\ref{fig:gmsb}
it is demonstrated how the different selections contribute to the
signal detection efficiency as a function of the NLSP lifetime.

None of the searches shows evidence for SUSY particle production. To interpret
the results, a detailed scan of the minimal GMSB parameter space is performed
with the gravitino mass fixed to 2 eV, corresponding to $\sqrt{F} \approx
100$~TeV, motivated by the requirement that the branching ratio of the
next-to-NLSP to the gravitino is small. If that is fulfilled, the
cross-sections and branching ratios do not depend on the gravitino mass. One
should note that $\sqrt{F}$ can be eliminated from the scan as all limits are
computed independent of the NLSP lifetime, and $\sqrt{F}$ has no significant
effect on other particle masses. 

"Model independent" cross-section limits are derived for each topology as a
function of the NLSP lifetime. For direct NLSP decays, this is done by  taking the
worst limit for a given NLSP mass from the generated GMSB parameter scan
points. For cascade channels, the cross-section evolution is  assumed to be
$\beta/s$ for spin-1/2 and $\beta^3/s$  for scalar SUSY particles,
respectively, and the highest bound for all intermediate particle masses is
retained. The maximum limit valid for all lifetimes is then quoted as the
"lifetime independent" cross-section limit. In the neutralino NLSP scenario
this is typically better than 0.04 pb for direct NLSP production, 0.1 pb for
selectron and smuon production, 0.2 pb for stau production and 0.3 pb for
chargino production. In the stau and slepton co-NLSP scenarios, the limit on
direct NLSP production is 0.05 pb for smuons, 0.1 pb for selectrons and
staus. For the cascade decays the bounds are typically better than 0.1 pb for
neutralino, 0.2 for chargino and in the stau NLSP scenario 0.4 for selectron
and smuon production.

The cross-section limits can be turned into constraints on the NLSP
mass. For sleptons, the lowest mass limits are found for very short
lifetimes, except for selectrons, shown in Figure~\ref{fig:gmsb}, where 
searches using d$E$/d$x$ measurements loose efficiency for particles with
momenta around 65 GeV. The lifetime independent limits are 
$m_{\tilde\mathrm{e}_\mathrm{R}} > 60.1$~GeV, $m_{\tilde\mu_\mathrm{R}} >
93.7$~GeV and  $m_{\tilde\tau_1} > 87.4$~GeV. The limit on the stau mass is the
same  in the stau and the slepton co-NLSP scenarios. In the slepton co-NLSP
scenario, the best limit can be used to derive a universal limit on the 
slepton masses $m_{\tilde\ell} = m_{\tilde\mu_\mathrm{R}} - m_\tau
> 91.9$~GeV, where by definition the mass differences between the different
slepton flavors are smaller than the lepton masses. For neutralino NLSP, no
lifetime independent NLSP mass limit can be set directly. For short lifetimes
($\tau < 10^{-9}$~s) a mass limit of 96.8 GeV is derived.

The GMSB parameter space can also be constrained by our results as shown in
Figure~\ref{fig:gmsb} for $N=1$, $M=1.01\cdot\Lambda$ and sign($\mu$)>0. The
universal SUSY mass scale is $\Lambda> 40, 27, 21, 17, 15$~TeV for messenger
indices $N=1, 2, 3, 4, 5$, independent of $M, \tan\beta$,
sign($\mu$) and the NLSP lifetime ($\sqrt{F}$). The constraints on $\Lambda$
imply lower limits on the neutralino mass in the neutralino NLSP scenario:
$m_{\tilde\chi_1^0} > 53.5$~GeV for N=1 and 
$m_{\tilde\chi_1^0} > 94.0$~GeV for N=5,
independent of the lifetime.

\enlargethispage{10 pt}

\end{document}